\documentclass[journal,twocolumn,10pt]{IEEEtran}
\usepackage{graphicx}
\usepackage{verbatim}
\usepackage{upgreek}
\usepackage{amssymb,amsmath}
\usepackage{color}
\usepackage{epstopdf}
\usepackage{bm}
\usepackage{cite}
\usepackage{array,color}
\usepackage{algorithm}
\usepackage{algorithmicx}
\usepackage{algpseudocode}
\usepackage{amsmath}
\usepackage{amsfonts}
\usepackage{amsthm}
\usepackage{graphics}
\usepackage{epsfig}
\usepackage{soul}
\usepackage{booktabs}
\usepackage{multirow}
\soulregister\cite7
\soulregister\ref7
\usepackage{subfigure}  
\usepackage{tabularx}
\usepackage{makecell}
\usepackage{graphicx}
\usepackage[font=small]{caption}

\usepackage{amsthm}
\newtheorem{remark}{Remark}

\usepackage{etoolbox} 
\makeatletter
\patchcmd{\@makecaption}
{\scshape}
{}
{}
{}
\makeatletter
\patchcmd{\@makecaption}
{\\}
{.\ }
{}
{}
\makeatother

\ifCLASSINFOpdf
\else
\fi

\title{Joint DNN Partitioning and Resource Allocation for 
Satellite--Terrestrial Collaborative Inference Systems}

\author{Wenyu Liang, Zesong Fei,~\IEEEmembership{Senior~Member,~IEEE}, Peng Liu, Xinyi Wang,~\IEEEmembership{Member,~IEEE}, and Ming Zeng

\thanks{The authors are with the School of Information and Electronics,
Beijing Institute of Technology, Beijing 100081, China
(e-mail: 13360736036@163.com, feizesong@bit.edu.cn,  bit\_peng\_liu@163.com, bit\_wangxy@163.com, and mzengzm@163.com.)}}
\begin{document}
\maketitle

\begin{abstract}
This letter investigates a satellite--terrestrial collaborative inference system for remote sensing information processing, where a ground station (GS) serves multiple low Earth orbit (LEO) satellites running heterogeneous deep neural network (DNN) inference tasks. By partitioning the DNN between the satellite and the GS, each satellite executes the front-end layers locally, transmits the resulting intermediate features to the GS, thereby offloading the remaining layers to the GS for inference completion. We jointly optimize DNN partitioning, satellite and GS computing resources, satellite--ground bandwidth allocation, and satellite transmit power to minimize the average task completion latency under per-satellite energy constraints. To address the coupling between discrete partitioning decisions and continuous communication-computing variables, we propose a two-layer optimization algorithm. The inner layer performs closed-form updates and efficient nested bisection searching to update the continuous resource variables, while the outer layer employs random-restart coordinate descent to refine the partitioning decisions. Simulation results show that the proposed algorithm effectively reduces task completion latency compared with conventional binary offloading.

\end{abstract}


\IEEEpeerreviewmaketitle
\section{Introduction}

Low Earth orbit (LEO) satellite constellations provide broad coverage and low propagation delay, enabling timely Earth observation for meteorological and environmental monitoring, resource exploration, and disaster response~\cite{leyva2020leo}. These applications increasingly rely on high-resolution remote sensing and deep neural network (DNN) inference, resulting in large data volumes, intensive computation, and considerable transmission and processing latency~\cite{diana2024review}. Conventional ground-centered processing requires the satellites to transmit raw data to the ground station (GS), incurring substantial satellite-to-GS downlink overhead, which is generally impractical under the short contact windows~\cite{denby2019orbital,leyva2023satellite}. Onboard edge computing alleviates this burden by processing data locally and transmitting only inference results~\cite{leyva2023satellite,chen2024energy}. However, fully onboard inference is constrained by limited power, computing, and thermal resources~\cite{delprete2025optimizing}. Binary offloading provides an alternative by restricting each DNN inference task to either full onboard execution or full GS execution. However, this coarse-grained strategy cannot exploit layer-wise variations in computational workload and intermediate feature size~\cite{eshratifar2021jointdnn}.

To address these limitations, DNN partitioning provides a fine-grained approach to collaborative inference across heterogeneous satellite--ground systems~\cite{eshratifar2021jointdnn,tian2026dynamic}. This approach exploits the layer-wise heterogeneity of DNN computation, where different layers exhibit diverse computational workloads and intermediate-feature sizes, and suitable bottleneck layers can generate compact features after moderate preceding computation. By partitioning the DNN at suitable bottleneck layers, the preceding layers are executed on the resource-constrained device, while the compact intermediate features are transmitted to a more capable server for executing the remaining workload. This avoids transmitting the raw input and executing the entire model locally, thereby enabling a finer tradeoff between computation and communication than binary offloading~\cite{kang2017neurosurgeon}.

There have been numerous efforts aiming at investigating computation offloading and collaborative inference in LEO satellite networks. In~\cite{tang2021computation}, the authors proposed a hybrid cloud-edge computing architecture for energy-efficient task offloading. Building on this architecture, the authors in~\cite{zhong2025joint} jointly optimized task offloading and resource allocation for heterogeneous tasks to maximize the aggregated task utility. Nevertheless, these studies considered only binary or workload-proportional offloading, without exploiting the layer-wise heterogeneity of DNN computation. To address this issue, the authors in~\cite{chen2023energy}  proposed an improved branch-and-bound method to optimize layer-wise DNN offloading, jointly reducing inference latency and onboard energy consumption. 

Beyond optimizing DNN partitioning alone, in~\cite{liu2025joint} and~\cite{liu2026toward}, the authors jointly optimized DNN partitioning and bandwidth resource allocation in terrestrial edge perception and ISCC networks, respectively, to reduce inference latency. However, the above studies~\cite{liu2025joint,liu2026toward} mainly focused on terrestrial networks, while DNN partitioning for satellite–terrestrial collaborative inference remains underexplored. Compared with terrestrial systems, satellite networks are constrained by limited onboard computing and energy resources, while multiple satellites compete for shared satellite–ground bandwidth and GS computing capacity under the short contact windows. These characteristics create a strong coupling between the DNN partition point and communication–computing resource allocation. Therefore, their joint optimization is essential for fully exploiting the complementary capabilities of onboard processing and ground-based computing, thereby reducing inference latency while satisfying satellite energy constraints.

Motivated by these observations, we investigate fine-grained DNN partitioning into multi-satellite–terrestrial collaborative inference. We jointly optimize the DNN partitioning vector, computing resources, satellite--ground bandwidth, and transmit power to minimize the average task completion latency subject to multi-dimensional resource and energy constraints. To solve the resulting mixed-integer nonconvex problem, we develop an efficient alternating algorithm using closed-form updates, nested bisection searching, and random-restart coordinate descent. Simulation results show that the proposed scheme outperforms binary offloading by better balancing onboard computation, feature transmission, and ground-side processing.

\section{System Model}
\begin{figure}[!t]
	\centering
	\includegraphics[width=2.7in]{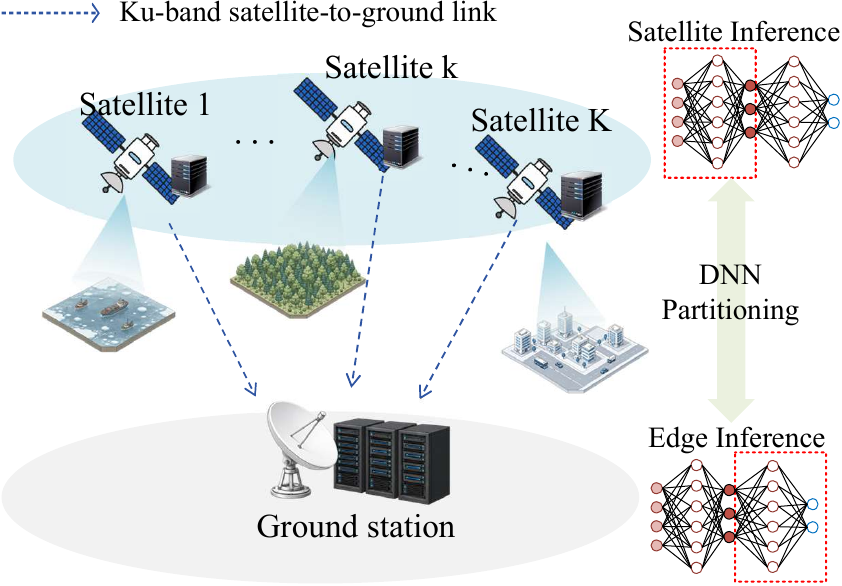}
	\caption{Satellite--Terrestrial Collaborative Inference Systems.}
	\vspace{-10pt}
	\label{fig:system_model}
\end{figure}

We consider a satellite--terrestrial collaborative edge computing system, as illustrated in Fig.~\ref{fig:system_model}, where a GS serves $K$ visible LEO satellites indexed by $k\in\mathcal{K}=\{1,\ldots,K\}$. The satellite--ground topology remains quasi-static, and multiple satellites access the GS through frequency-division multiple access. The satellites generate multiple DNN inference tasks, with task arrival rates and model architectures varying across satellites. Satellite $k$ is visible to the GS when its elevation angle $\theta_k$ satisfies $\theta_k\geq\theta_{\min}$. The distance between satellite $k$ and the GS is given by

{\small
	\setlength{\abovedisplayskip}{-8pt} 
	\setlength{\belowdisplayskip}{5pt} 
\begin{equation}
d_k=
\sqrt{(R_{\mathrm{e}}+H)^2-(R_{\mathrm{e}}\cos\theta_k)^2}
-R_{\mathrm{e}}\sin\theta_k ,
\end{equation}
}where $H$ and $R_{\mathrm{e}}$ denote the orbital altitude of the LEO satellites and the Earth radius, respectively.

The achievable transmission rate from satellite $k$ to the GS is modeled as

{\small
	\setlength{\abovedisplayskip}{-10pt} 
	\setlength{\belowdisplayskip}{5pt} 
\begin{equation}
r_k=B_k\log_2\left(1+\frac{p_k^{\mathrm{s}} h_k}{N_0B_k}\right),
\end{equation}
}where $B_k$, $p_k^{\mathrm{s}}$, and $N_0$ denote the allocated bandwidth, satellite transmit power, and noise power spectral density, respectively. The channel gain $h_k$ incorporates antenna gains, free-space path loss, elevation-dependent atmospheric and rain attenuation, and shadowed-Rician fading.

For each layer $l\in\mathcal{L}_k$, we denote its input and output tensor shapes by
$\mathbf{s}_{k,l}^{\rm in}
=(C_{k,l}^{\rm in},H_{k,l}^{\rm in},W_{k,l}^{\rm in})$
and
$\mathbf{s}_{k,l}^{\rm out}
=(C_{k,l}^{\rm out},H_{k,l}^{\rm out},W_{k,l}^{\rm out})$,
respectively. Accordingly, the number of output elements of layer $l$ is given by

{\small
	\setlength{\abovedisplayskip}{-8pt} 
	\setlength{\belowdisplayskip}{5pt} 
\begin{equation}
N_{k,l}
=C_{k,l}^{\rm out}H_{k,l}^{\rm out}W_{k,l}^{\rm out}.
\end{equation}
}

Let $\omega_{k,l}$ denote the operator type of layer $l$, and define
$\chi_{k,l}=C_{k,l}^{\rm in}K_{k,l}^{\mathrm{h}}K_{k,l}^{\mathrm{w}}/g_{k,l}$,
where $K_{k,l}^{\mathrm{h}}$, $K_{k,l}^{\mathrm{w}}$, and $g_{k,l}$ denote the convolution kernel dimensions and the number of groups, respectively. The fine-grained arithmetic workload of layer $l$, measured in floating-point operations (FLOPs), is modeled as\cite{li2023throughput}

{\small
	\setlength{\abovedisplayskip}{-8pt} 
	\setlength{\belowdisplayskip}{5pt} 
\begin{equation}
F_k^{\rm DNN}(l)=
\begin{cases}
(2\chi_{k,l}+b_{k,l})N_{k,l},
& \omega_{k,l}=\mathrm{Convolution},\\
4N_{k,l},
& \omega_{k,l}=\mathrm{Batch\ Normalization},\\
N_{k,l},
& \omega_{k,l}=\mathrm{Activation},\\
(K_{k,l}^{\mathrm{h}}K_{k,l}^{\mathrm{w}}-1)N_{k,l},
& \omega_{k,l}=\mathrm{Max\ Pooling},\\
\bar{K}_{k,l}^{\mathrm{h}}\bar{K}_{k,l}^{\mathrm{w}}N_{k,l},
& \omega_{k,l}=\mathrm{Average\ Pooling},\\
N_{k,l},
& \omega_{k,l}=\mathrm{Element\text{-}wise\ Addition},\\
2d_{k,l}^{\rm in}d_{k,l}^{\rm out}
+b_{k,l}d_{k,l}^{\rm out},
& \omega_{k,l}=\mathrm{Fully\ Connected},\\
0,
& \omega_{k,l}\in\mathcal{Z}.
\end{cases}
\end{equation}
}

Here, $b_{k,l}\in\{0,1\}$ indicates whether a bias term is used,
$d_{k,l}^{\rm in}$ and $d_{k,l}^{\rm out}$ denote the input and output
dimensions of a fully connected layer, and
$\bar{K}_{k,l}^{\mathrm{h}}\bar{K}_{k,l}^{\mathrm{w}}$ denotes the average number of input elements contributing to each adaptive-average-pooling output. The set $\mathcal{Z}$ denotes non-computational operators, including the virtual input layer, Flatten, and Dropout, whose arithmetic workloads are negligible during inference. Under $q$-bit quantization, the output feature size after layer $l$ is ${\rm Data}_k(l)=qN_{k,l}$. Accordingly, each DNN task is characterized by the layer-wise workload profile
$\{F_k^{\rm DNN}(l)\}_{l=0}^{L_k}$ and feature-size profile
$\{{\rm Data}_k(l)\}_{l=0}^{L_k}$.

For satellite $k$, the resulting candidate set is denoted by $\mathcal{I}_k=\{i_{k,0},i_{k,1},\ldots,i_{k,M_k-1}\}$, where $M_k$ is the number of candidates and $\delta(i_k)$ maps $i_k\in\mathcal{I}_k$ to the corresponding DNN layer index. In particular, $\delta(i_{k,0})=0$ and $\delta(i_{k,M_k-1})=L_k$ represent full GS execution and full onboard execution, respectively. Given partition point $i_k$, the resulting satellite--ground transmission data size is $D_k={\rm Data}_k\bigl(\delta(i_k)\bigr)$, while the satellite-side and ground-side workloads are given by

{\small
	\setlength{\abovedisplayskip}{-5pt} 
	\setlength{\belowdisplayskip}{3pt} 
\begin{equation}
\begin{aligned}
F_k^{\mathrm{s}}
&=\sum_{l=0}^{\delta(i_k)}F_k^{\rm DNN}(l),
&\qquad
F_k^{\mathrm{g}}
&=\sum_{l=\delta(i_k)+1}^{L_k}F_k^{\rm DNN}(l).
\end{aligned}
\end{equation}
}

Let $A_k$ denote the number of tasks generated by satellite $k$ in a time slot. Tasks arriving in the same slot are processed as a batch using the same partitioning decision. With $c$ denoting the speed of light, the transmission latency and propagation latency are respectively expressed as
{\small
\setlength{\abovedisplayskip}{5pt}
\setlength{\belowdisplayskip}{5pt}
\begin{equation}
\begin{aligned}
T_k^{\rm trans}
&= \frac{A_kD_k}{r_k},
&\qquad
T_k^{\rm pro}
&= \frac{d_k}{c}.
\end{aligned}
\end{equation}
}

Let $f_k^{\mathrm{s}}$ and $f_k^{\mathrm{g}}$ denote the onboard and GS computing resources allocated to satellite $k$, respectively. The satellite-side and ground-side computation latencies are given by

{\small
	\setlength{\abovedisplayskip}{-8pt} 
	\setlength{\belowdisplayskip}{5pt} 
\begin{equation}
\begin{aligned}
T_k^{\mathrm{s,com}}
&= \frac{A_kF_k^{\mathrm{s}}}{f_k^{\mathrm{s}}},
&\qquad
T_k^{\mathrm{g,com}}
&= \frac{A_kF_k^{\mathrm{g}}}{f_k^{\mathrm{g}}}.
\end{aligned}
\end{equation}
}

Accordingly, the end-to-end task completion latency of satellite $k$ is

{\small
	\setlength{\abovedisplayskip}{-8pt} 
	\setlength{\belowdisplayskip}{5pt} 
\begin{equation}
T_k =
T_k^{\rm trans}
+ T_k^{\rm pro}
+ T_k^{\mathrm{s,com}}
+ T_k^{\mathrm{g,com}}.
\end{equation}
}

The energy consumption of satellite $k$ mainly arises from onboard computation and data transmission. Since satellite static energy is independent of the partitioning decision and the GS has sufficient energy supply, the static and ground-side energy consumption are omitted. Thus, the energy consumption of satellite $k$ is given by

{\small
	\setlength{\abovedisplayskip}{-8pt} 
	\setlength{\belowdisplayskip}{5pt} 
\begin{equation}
E_k
=
p_k^{\mathrm{c}}T_k^{\mathrm{s,com}}
+
p_k^{\mathrm{s}}T_k^{\rm trans},
\end{equation}
}where $p_k^{\mathrm{c}}=\kappa(f_k^{\mathrm{s}})^3$ denotes the dynamic power of the onboard processor, with $\kappa$ being the effective switched capacitance coefficient. 

\section{Problem Formulation and Algorithm Design}
\subsection{Problem Formulation}
In this section, we aim to jointly optimize the DNN partitioning vector $\mathbf{i}=[i_1,\ldots,i_K]^{\mathrm{T}}$, satellite computing resource allocation vector $\mathbf{f}^{\mathrm{s}}=[f_1^{\mathrm{s}},\ldots,f_K^{\mathrm{s}}]^{\mathrm{T}}$, GS computing resource allocation vector $\mathbf{f}^{\mathrm{g}}=[f_1^{\mathrm{g}},\ldots,f_K^{\mathrm{g}}]^{\mathrm{T}}$, satellite--ground bandwidth allocation vector $\mathbf{b}=[B_1,\ldots,B_K]^{\mathrm{T}}$, and satellite transmit power control vector $\mathbf{p}^{\mathrm{s}}=[p_1^{\mathrm{s}},\ldots,p_K^{\mathrm{s}}]^{\mathrm{T}}$ to minimize the average task completion latency across all connected satellites, subject to the computing-resource, bandwidth, transmit-power, and satellite-energy constraints. Specifically, the optimization problem is formulated as

{\small
\setlength{\abovedisplayskip}{-8pt}
\setlength{\belowdisplayskip}{5pt}
\begin{subequations}\label{P1}
\begin{align}
\min_{\substack{
\mathbf{i},\,\mathbf{f}^{\mathrm{s}},\,\mathbf{f}^{\mathrm{g}},\\
\mathbf{b},\,\mathbf{p}^{\mathrm{s}}
}}
\quad
& \frac{1}{K}\sum_{k=1}^{K}T_k
\tag{\theparentequation}
\label{P1obj}\\
\mathrm{s.t.}\quad
& i_k\in\mathcal{I}_k,
\quad \forall k\in\mathcal{K},
\label{P1a}\\
& 0\leq f_k^{\mathrm{s}}\leq f_{\max}^{\mathrm{s}},
\quad \forall k\in\mathcal{K},
\label{P1b}\\
& \sum_{k=1}^{K}f_k^{\mathrm{g}}\leq f_{\max}^{\mathrm{g}},
\quad f_k^{\mathrm{g}}\geq 0,
\quad \forall k\in\mathcal{K},
\label{P1c}\\
& \sum_{k=1}^{K}B_k\leq B_{\max},
\quad B_k\geq 0,
\quad \forall k\in\mathcal{K},
\label{P1d}\\
& 0\leq p_k^{\mathrm{s}}\leq p_{\max}^{\mathrm{s}},
\quad \forall k\in\mathcal{K},
\label{P1e}\\
& E_k\leq E_{\max},
\quad \forall k\in\mathcal{K}.
\label{P1f}
\end{align}
\end{subequations}
}where \eqref{P1a} restricts the partition point of satellite $k$ to the predefined candidate set $\mathcal{I}_k$. Constraints \eqref{P1b} and \eqref{P1c} limit the computing resources of each satellite and the GS to $f_{\max}^{\mathrm{s}}$ and $f_{\max}^{\mathrm{g}}$, respectively. Constraint \eqref{P1d} ensures that the total allocated satellite-to-ground bandwidth does not exceed $B_{\max}$. Constraint \eqref{P1e} specifies the maximum transmit power $p_{\max}^{\mathrm{s}}$ of each satellite. Finally, constraint \eqref{P1f} guarantees that the energy consumption of each satellite remains within the prescribed energy budget $E_{\max}$.

\subsection{Algorithm Design}
We observe that problem \eqref{P1} is challenging to solve due to the coexistence of discrete DNN partitioning decisions and continuous resource allocation variables, the coupling among computation latency, transmission latency, and energy consumption, as well as the nonlinear coupling between bandwidth and transmit power. To address these challenges, we propose a two-layer optimization algorithm as described below.

\subsubsection{Inner-Layer Continuous Resource Allocation Problem}
For a given DNN partitioning vector $\mathbf{i}$, ${F_k^\mathrm{s},F_k^\mathrm{g},D_k}$ are fixed, and problem \eqref{P1} reduces to the following continuous resource allocation problem:

{\small
\setlength{\abovedisplayskip}{-8pt}
\setlength{\belowdisplayskip}{5pt}
\begin{equation}\label{P2}
\begin{aligned}
\makebox[1.1cm][l]{$\displaystyle
\min_{\substack{
\mathbf{f}^{\mathrm{s}},\,\mathbf{f}^{\mathrm{g}},
\mathbf{b},\,\mathbf{p}^{\mathrm{s}}
}}$}
&\quad
\frac{1}{K}\sum_{k=1}^{K}T_k
\\[-1pt]
\makebox[0.6cm][l]{$\displaystyle \mathrm{s.t.}$}
&\quad
\eqref{P1b}\text{--}\eqref{P1f}.
\end{aligned}
\end{equation}
}

Due to the nonlinear coupling among computation, communication, and energy variables, we introduce the Lagrange multipliers $\nu\geq0$, $\mu\geq0$, and $\lambda_k\geq0$ for the GS computing, total bandwidth, and satellite energy constraints, respectively, yielding the following partial Lagrangian:

{\small
	\setlength{\abovedisplayskip}{-8pt} 
	\setlength{\belowdisplayskip}{2pt} 
\begin{align}
\mathcal{L}
=&
\frac{1}{K}\sum_{k=1}^{K}
\left(
\frac{A_kD_k}{r_k}
+\frac{d_k}{c}
+\frac{A_kF_k^{s}}{f_k^{s}}
+\frac{A_kF_k^{g}}{f_k^{g}}
\right)
\notag\\
&+
\nu\left(\sum_{k=1}^{K}f_k^{g}-f_{\max}^{\mathrm{g}}\right)
+
\mu\left(\sum_{k=1}^{K}B_k-B_{\max}\right)
\notag\\
&+
\sum_{k=1}^{K}\lambda_k
\left(
\kappa A_kF_k^{s}(f_k^{s})^2
+
p_k^{s}\frac{A_kD_k}{r_k}
-
E_{\max}
\right).
\label{Lag}
\end{align}
}

The individual box constraints on $f_k^{s}$ and $p_k^{s}$ are enforced during the primal-variable updates. For fixed dual variables $(\nu,\mu,\boldsymbol{\lambda})$, the Lagrangian in \eqref{Lag} decomposes across satellites, allowing the resource variables of each satellite to be updated independently.

The optimal GS computing resource allocation is obtained from the
Karush--Kuhn--Tucker (KKT) conditions~\cite{boyd2004convex}, after assigning
$f_k^{\mathrm{g}*}=0$ to satellites satisfying
$A_kF_k^{\mathrm{g}}=0$, as

{\small
	\setlength{\abovedisplayskip}{-8pt} 
	\setlength{\belowdisplayskip}{5pt} 
\begin{equation}\label{fg_star}
f_k^{\mathrm{g}*}
=
\frac{\sqrt{A_kF_k^{g}}}
{\sum_{j=1}^{K}\sqrt{A_jF_j^{g}}}
f_{\max}^{\mathrm{g}}.
\end{equation}
}

\begin{remark}
It can be observed from \eqref{fg_star} that the GS computing capacity is allocated in proportion to the square root of each satellite’s ground-side computational workload.
\end{remark}

Moreover, the optimal satellite computing frequency is obtained from
the KKT conditions associated with \eqref{Lag}. For satellites with
$A_kF_k^{\mathrm{s}}=0$, we set $f_k^{\mathrm{s}*}=0$. Otherwise,
{\small
\setlength{\abovedisplayskip}{3pt}
\setlength{\belowdisplayskip}{2pt}
\begin{equation}\label{fs_star}
f_k^{\mathrm{s}*}
=
\begin{cases}
f_{\max}^{\mathrm{s}}, & \lambda_k=0,\\[1mm]
\min\left\{
f_{\max}^{\mathrm{s}},
\left(2K\lambda_k\kappa\right)^{-\frac{1}{3}}
\right\}, & \lambda_k>0.
\end{cases}
\end{equation}
}

Specifically, when the energy constraint is nonbinding, $\lambda_k=0$ and the satellite operates at $f_{\max}^{\mathrm{s}}$ to minimize latency; when it becomes binding, $\lambda_k>0$ and the computing frequency is reduced to satisfy the energy budget.

Next, we jointly optimize the bandwidth and transmit power. When $D_k=0$, no intermediate features are transmitted, and hence $B_k^{*}=p_k^{\mathrm{s}*}=0$. For $D_k>0$, we introduce $x_k=p_k^{\mathrm{s}}h_k/(N_0B_k)$, yielding $B_k=p_k^{\mathrm{s}}h_k/(N_0x_k)$ and $r_k=[p_k^{\mathrm{s}}h_k/(N_0x_k)]\log_2(1+x_k)$. Substituting these expressions into \eqref{Lag}, the optimal transmit power for a given $x_k$ is obtained as

{\small
	\setlength{\abovedisplayskip}{-5pt}
	\setlength{\belowdisplayskip}{-5pt}
\begin{equation}\label{p_x}
p_k^{\mathrm{s}*}(x_k)=
\begin{cases}
\displaystyle
\min\left\{
p_{\max}^{\mathrm{s}},
\sqrt{
\frac{A_kD_k(N_0x_k)^2}
{K\mu h_k^2\log_2(1+x_k)}
}
\right\},
& \mu>0,\\[4mm]
p_{\max}^{\mathrm{s}},
& \mu=0.
\end{cases}
\end{equation}
}

When $\mu=0$, the transmit-power update reduces to $p_k^{\mathrm{s}*}=p_{\max}^{\mathrm{s}}$. However, if $D_k>0$ for at least one satellite, the total-bandwidth constraint is active at the optimum, and the KKT stationarity condition further implies $\mu^*>0$. Hence, the scalar optimization of $x_k$ relevant to the optimal solution is considered for $\mu>0$. To update $x_k$, we define its corresponding objective function as

{\small
	\setlength{\abovedisplayskip}{0pt}
	\setlength{\belowdisplayskip}{5pt}
\begin{align}
\Phi_k(x_k)
=&
\frac{A_kD_kN_0x_k}
{Kh_k p_k^{\mathrm{s}*}(x_k)\log_2(1+x_k)}
+
\lambda_k
\frac{A_kD_kN_0x_k}
{h_k\log_2(1+x_k)}
\notag\\
&+
\mu
\frac{p_k^{\mathrm{s}*}(x_k)h_k}{N_0x_k}.
\label{Phi_x}
\end{align}
}

For $\mu>0$, $\Phi_k(x_k)$ is strictly unimodal over $x_k>0$, as its derivative changes sign at most once from negative to positive across the power-control regimes in \eqref{p_x}. Hence, the unique minimizer $x_k^{*}$ can be obtained by solving $\partial\Phi_k(x_k)/\partial x_k=0$ via bisection search. Subsequently, $B_k^{*}=p_k^{\mathrm{s}*}(x_k^{*})h_k/(N_0x_k^{*})$.

The dual variables are determined by enforcing the complementary slackness conditions, with the corresponding roots obtained via nested bisection search. For each $\mu$, the inner bisection search updates $\{\lambda_k\}_{k=1}^{K}$ to satisfy the energy constraints, while the outer bisection search updates $\mu$ to satisfy the total-bandwidth constraint.

\subsubsection{Outer-Layer DNN Partitioning Problem}
In the outer layer, we determine the DNN partitioning vector $\mathbf{i}$, which specifies the workloads and transmitted feature size for the inner-layer problem \eqref{P2}. By solving \eqref{P2}, the optimal computing, bandwidth, and power variables are obtained. Let $\bar{T}^{*}(\mathbf{i})$ denote the corresponding minimum average latency. The outer-layer problem is then formulated as

{\small
	\setlength{\abovedisplayskip}{-3pt} 
	\setlength{\belowdisplayskip}{2pt} 
\setlength{\jot}{0pt}
\begin{subequations}\label{P3}
\begin{align}
\min_{\mathbf{i}}\quad
& \bar{T}^{*}(\mathbf{i})
\notag\\
\mathrm{s.t.}\quad
& i_k\in\mathcal{I}_k,\quad \forall k\in\mathcal{K}.
\label{P3a}
\end{align}
\end{subequations}
}

Since the solution space size $\prod_{k=1}^{K}|\mathcal{I}_k|$ makes exhaustive search intractable, we employ a random-restart coordinate descent method to efficiently approximate the global optimum.

The algorithm updates one coordinate at a time. For satellite $k$, each candidate $\widetilde{i}_k\in\mathcal{I}_k$ forms the trial vector $\mathbf{i}^{\mathrm{test}}(\widetilde{i}_k)=[i_1,\ldots,i_{k-1},\widetilde{i}_k,i_{k+1},\ldots,i_K]^{\mathrm{T}}$. The $k$-th coordinate is then updated as

{\small
	\setlength{\abovedisplayskip}{-10pt} 
	\setlength{\belowdisplayskip}{5pt} 
\begin{equation}\label{update_rule}
i_k
\leftarrow
\arg\min_{\widetilde{i}_k\in\mathcal{I}_k}
\bar{T}^{*}
\left(
\mathbf{i}^{\mathrm{test}}(\widetilde{i}_k)
\right).
\end{equation}
}

This coordinate descent cycles over all $k\in\mathcal{K}$ and terminates when $I_{\max}^{\mathrm{out}}$ is reached or a full sweep yields no strict objective reduction, indicating convergence to a local optimum. To reduce sensitivity to local minima, multiple restarts are performed from $\mathcal{S}_{\mathrm{init}}$, which contains predefined representative configurations and randomly sampled feasible partitions.

\renewcommand{\algorithmicrequire}{\textbf{Input:}}
\renewcommand{\algorithmicensure}{\textbf{Output:}}

\begin{algorithm}[t]
\caption{Overall Algorithm for Solving Problem \eqref{P1}}
\label{alg_LPORA}
\begin{algorithmic}[1]
\Require Outer layer:  system parameters and candidate partition sets
$\{\mathcal{I}_k\}_{k=1}^{K}$, initial set
$\mathcal{S}_{\mathrm{init}}$, restart number $R$, and
$I_{\max}^{\mathrm{out}}$;
\Statex \hspace{\algorithmicindent}
Inner layer: system parameters and bisection searching iterations
$N_{\mu}$, $N_{\lambda}$, and $N_x$.

\State Initialize $\bar{T}^{*}=+\infty$.

\For{$r=1,\ldots,R$}
    \State Select an initial partitioning vector
    $\mathbf{i}\in\mathcal{S}_{\mathrm{init}}$.
    \State Set $m=0$.
    \Repeat
        \State Generate trial partitioning vectors
        $\mathbf{i}^{\mathrm{test}}(\widetilde{i}_k)$.
        \State Determine $\{F_k^{s},F_k^{g},D_k\}_{k=1}^{K}$
        for each trial vector.
        \State Optimize $\mathbf{f}^{\mathrm{g*}}$ ,
        $\mathbf{f}^{\mathrm{s*}}$ via the KKT updates
        \eqref{fg_star},\eqref{fs_star}.
        \State Optimize $\{x_k^*\}_{k=1}^{K}$
        via scalar bisection using \eqref{Phi_x}.
        \State Obtain $\boldsymbol{p}^{s*}$ and
        $\boldsymbol{b}^{*}$ via \eqref{p_x}.
        \State Update $\{\lambda_k\}_{k=1}^{K}$ and $\mu$ via nested bisection searching based on the complementary slackness conditions.
        \State Evaluate
        $\bar{T}^{*}(\mathbf{i}^{\mathrm{test}})$.
        \State Update $\mathbf{i}$ coordinate-wise
        via \eqref{update_rule}.
        \State Retain the corresponding resource allocation.
        \State Set $m=m+1$.
    \Until{the partitioning vector converges or
    $m=I_{\max}^{\mathrm{out}}$.}

    \If{$\bar{T}^{*}(\mathbf{i})<\bar{T}^{*}$}
        \State Update $\mathbf{i}^{*}=\mathbf{i}$ and
        $\bar{T}^{*}=\bar{T}^{*}(\mathbf{i})$.
        \State Store $\mathbf{f}^{\mathrm{s*}}$,
        $\mathbf{f}^{\mathrm{g*}}$,
        $\mathbf{b}^{*}$, and
        $\mathbf{p}^{\mathrm{s*}}$.
    \EndIf
\EndFor

\Ensure $\mathbf{i}^{*}$,
$\mathbf{f}^{\mathrm{s*}}$,
$\mathbf{f}^{\mathrm{g*}}$,
$\mathbf{b}^{*}$,
$\mathbf{p}^{\mathrm{s*}}$,
and $\bar{T}^{*}$.
\end{algorithmic}
\end{algorithm}

\subsubsection{Overall Algorithm and Complexity}
The complete two-layer optimization framework is summarized in Algorithm~\ref{alg_LPORA}. Let $M_{\max}=\max_k|\mathcal{I}_k|$, and let $N_{\mu}$, $N_{\lambda}$, and $N_x$ denote the corresponding bisection searching iterations. The inner resource allocation for each trial vector requires $\mathcal{O}(KN_{\mu}N_{\lambda}N_x)$ operations, while each coordinate-descent iteration evaluates at most $\mathcal{O}(KM_{\max})$ trials. Hence, the overall worst-case complexity is $\mathcal{O}(RI_{\max}^{\mathrm{out}}K^2M_{\max}N_{\mu}N_{\lambda}N_x)$.

\section{Simulation Results}
In this section, we evaluate the performance of the proposed algorithm through Monte Carlo simulations. We consider a satellite--terrestrial collaborative inference system comprising one GS and $K=8$ LEO satellites at an orbital altitude of $H=600$ km. The satellite--ground channels account for free-space path loss, random shadow fading, and link perturbations. The carrier frequency, total system bandwidth, transmit antenna gain, receive antenna gain, and noise power spectral density are set to $f_{\mathrm{c}}=12$ GHz, $B_{\max}=80$ MHz, $G_{\mathrm{t}}=25$ dB, $G_{\mathrm{r}}=35$ dB, and $N_0=-174$ dBm/Hz, respectively. Unless otherwise specified, $f_{\max}^g=1.8\times10^{12}$ FLOPs/s, $f_{\max}^s=2.5\times10^{10}$ FLOPs/s, $p_{\max}^s=20$ W, and $\kappa=6\times10^{-32}$. The intermediate features are quantized to $q=8$ bits. To capture heterogeneous computation demands, half of the satellites are assigned AlexNet-based inference tasks, while the remaining satellites execute ResNet18-based inference tasks.

For comparison, we evaluate the proposed algorithm against five benchmarks. \emph{AO-GS} employs alternating optimization with Gauss--Seidel-type block updates for continuous resource allocation~\cite{razaviyayn2013unified}. \emph{Random partitioning (RP)} randomly selects the DNN partitioning vector. \emph{Equal bandwidth} and \emph{fixed transmit power} adopt uniform bandwidth allocation and identical transmit power, respectively. \emph{Binary offloading} restricts each task to either full onboard execution or full GS execution. Except for the specified differences, all baseline schemes use the same settings as the proposed algorithm.

\begin{figure}[!t]
\centering
\setlength{\subfigcapskip}{-5pt} 

\subfigure[]{%
    \includegraphics[
        width=0.49\columnwidth,
        trim=2mm 1mm 2mm 1mm,
        clip
    ]{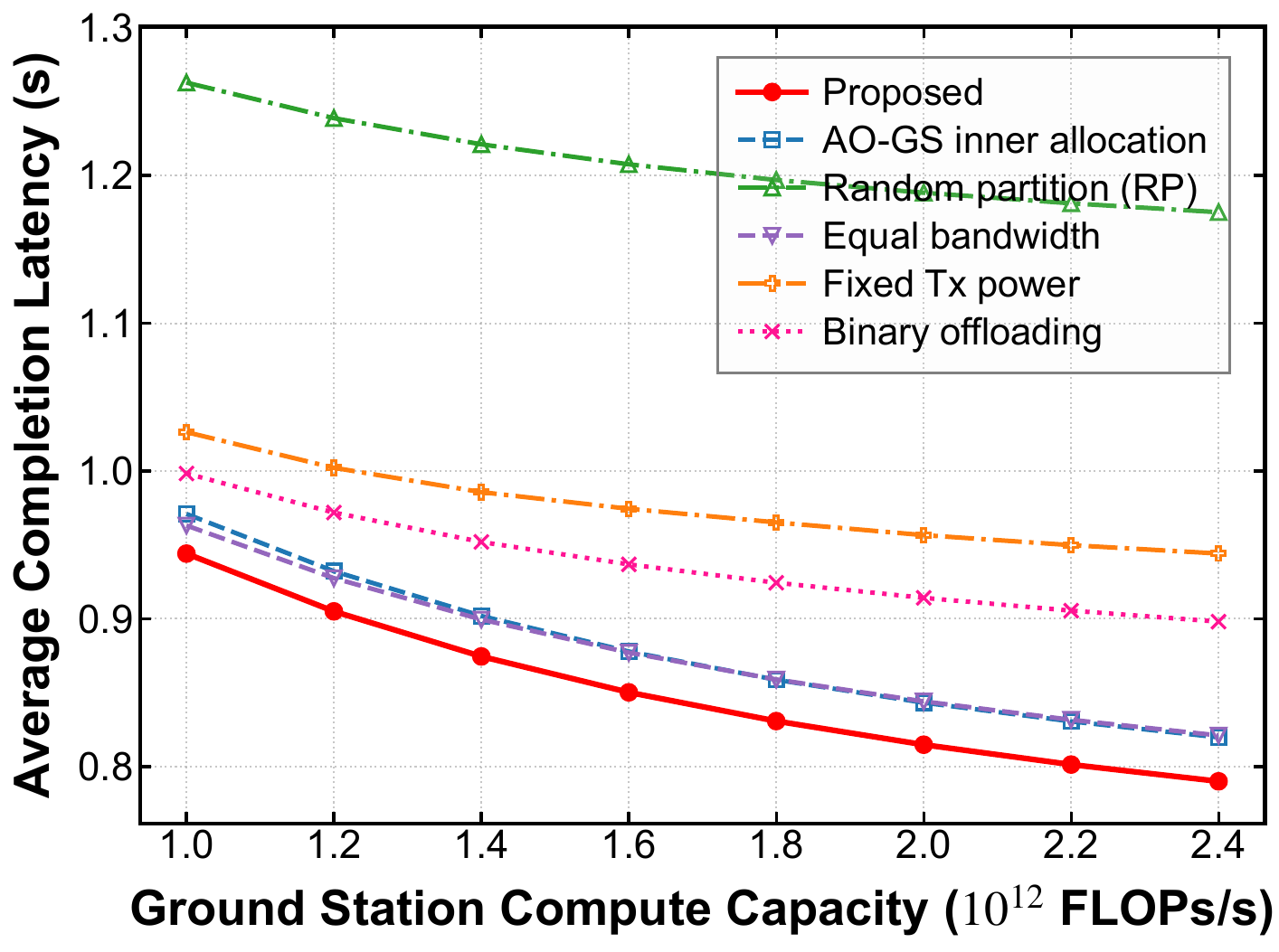}%
    \label{fig:gs_compute}%
}%
\hspace{0.01\columnwidth}%
\subfigure[]{%
    \includegraphics[
        width=0.49\columnwidth,
        trim=2mm 1mm 2mm 1mm,
        clip
    ]{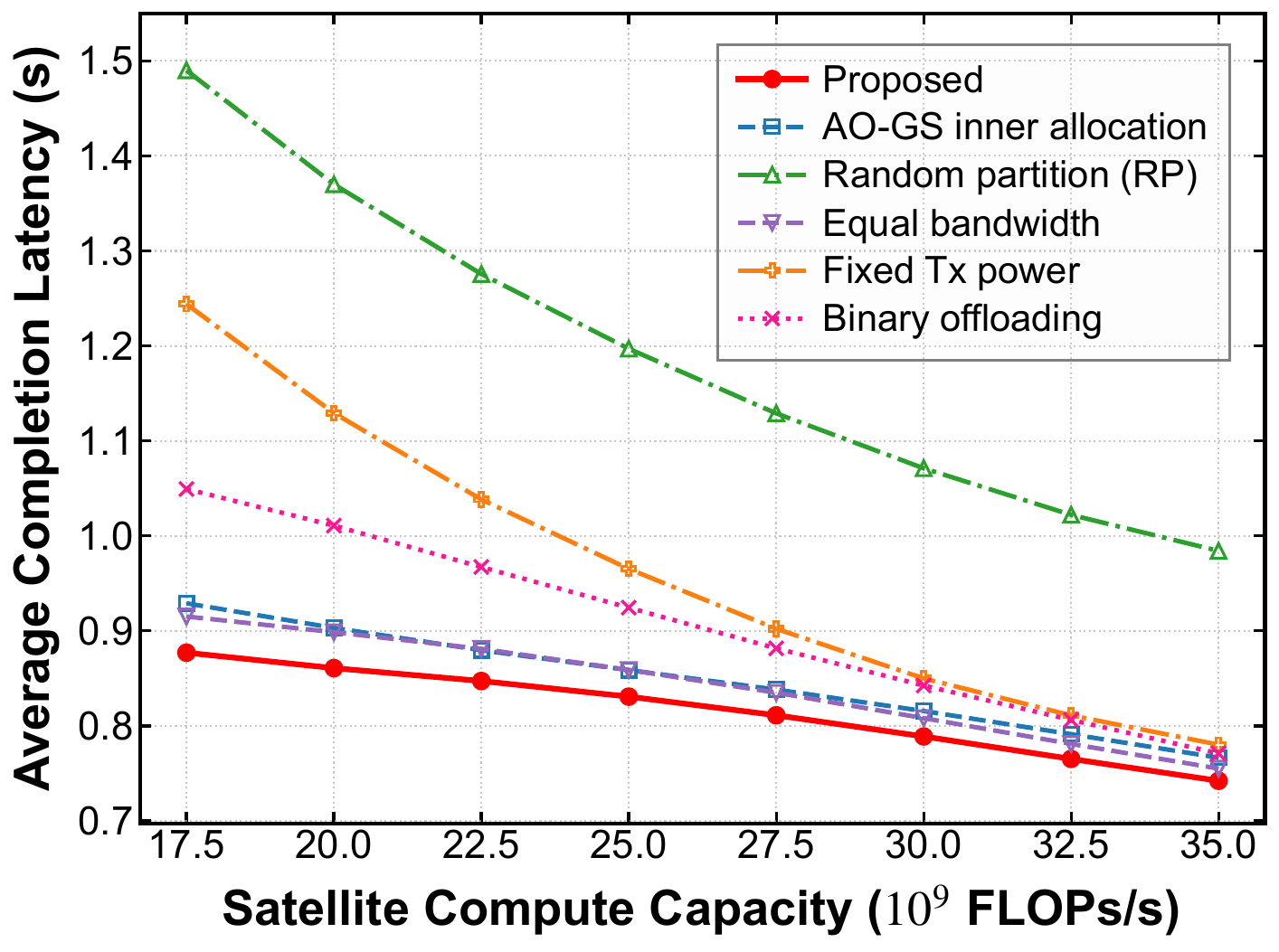}%
    \label{fig:sat_compute}%
}%

\caption{Average task completion latency versus
(a) GS computing capacity and
(b) satellite computing capacity.}
\label{fig:compute}
\vspace{-10pt}
\end{figure}

Fig.~\ref{fig:compute}(a) shows the average task completion latency versus the GS computing capacity. The latency of all schemes decreases as the GS computing capacity increases, while the advantage of the proposed algorithm over the partitioning baselines becomes more pronounced. This is because the proposed algorithm adapts the DNN partitioning vector to offload appropriate DNN layers to the GS, thereby utilizing the additional GS resources without excessive intermediate-feature transmission. By contrast, random partitioning and binary offloading provide less flexibility in balancing computation and transmission. Fig.~\ref{fig:compute}(b) shows the average task completion latency versus the satellite computing capacity. Increasing the satellite computing capacity alleviates the onboard computation bottleneck and reduces latency. The performance gaps consequently narrow because DNN partitioning becomes less critical when onboard resources are abundant. Nevertheless, the proposed algorithm consistently achieves the lowest latency by adapting the partitioning decisions to the dominant system bottleneck.

\begin{figure}[!t]
\centering
\setlength{\subfigcapskip}{-5pt} 
\subfigure[]{%
    \includegraphics[
        width=0.49\columnwidth,
        trim=2mm 1mm 2mm 1mm,
        clip
    ]{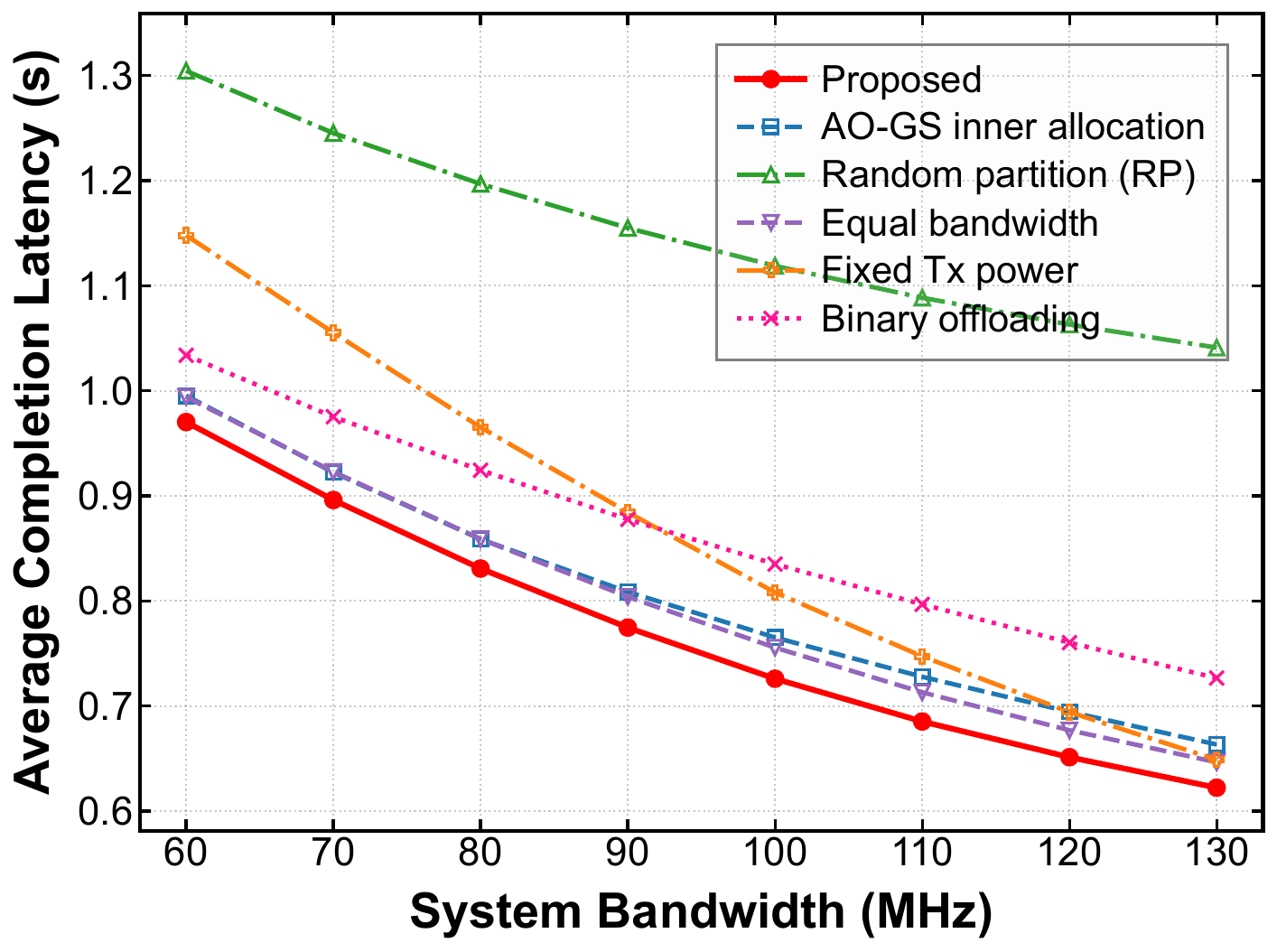}%
    \label{fig:bandwidth}%
}%
\hspace{0.01\columnwidth}%
\subfigure[]{%
    \includegraphics[
        width=0.49\columnwidth,
        trim=2mm 1mm 2mm 1mm,
        clip
    ]{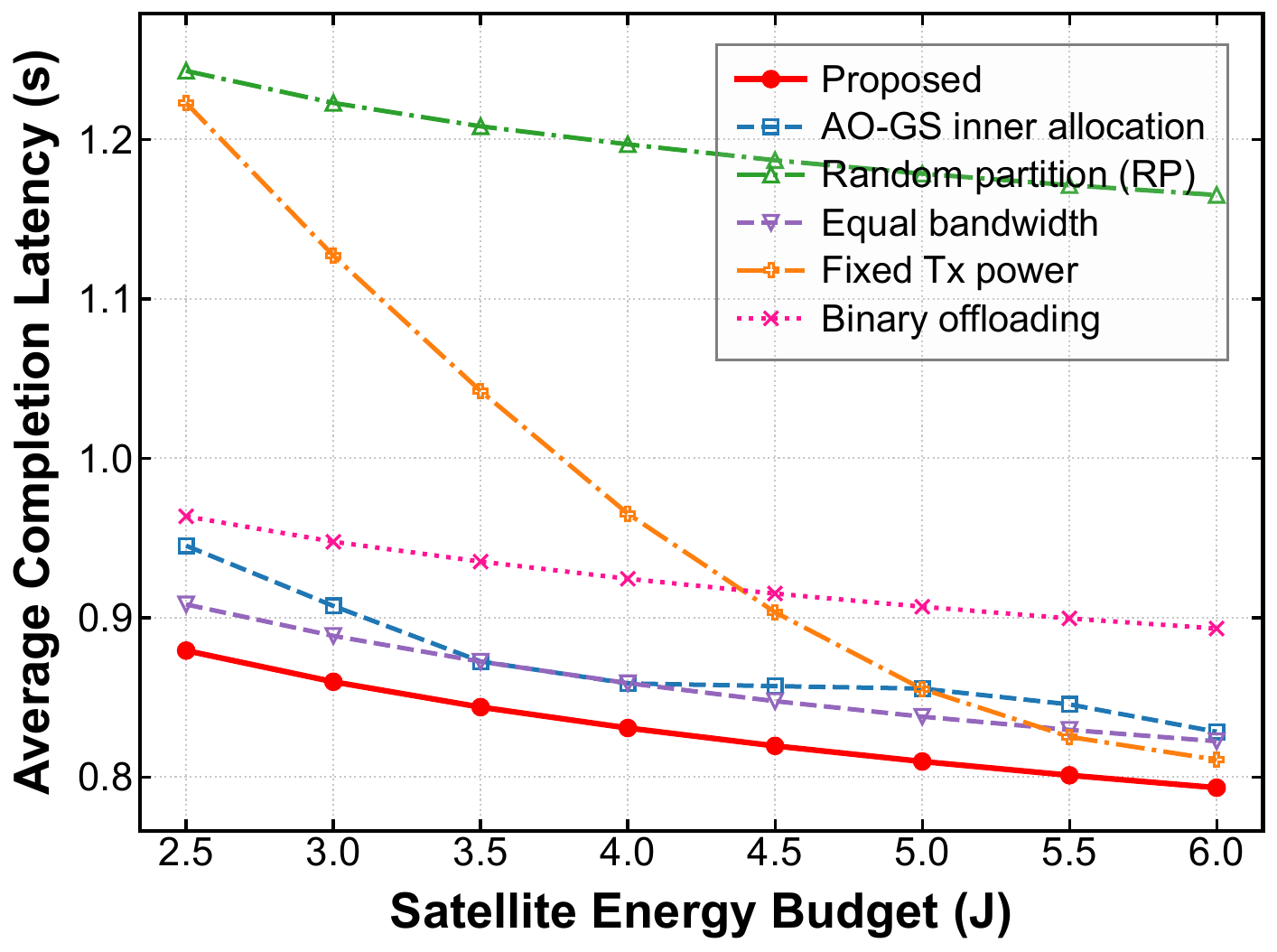}%
    \label{fig:energy}%
}%
\caption{Average task completion latency versus
(a) system bandwidth and
(b) satellite energy budget.}
\label{fig:communication}
\vspace{-10pt}
\end{figure}

Fig.~\ref{fig:communication}(a) shows the average task completion latency under different system bandwidths. We observe that increasing the bandwidth improves the transmission rate and reduces the overall latency, while computation gradually becomes the dominant latency component. The proposed algorithm consistently outperforms equal bandwidth by jointly adapting the bandwidth allocation and DNN partitioning vector to heterogeneous feature sizes and channel conditions, thereby enabling more appropriate GS-side execution without excessive transmission overhead. Fig.~\ref{fig:communication}(b) shows the average task completion latency under different satellite energy budgets. A larger energy budget supports higher computing frequencies and more flexible transmit-power allocation, reducing both computation and transmission latency. The advantage of the proposed algorithm over fixed transmit power is more evident under stringent energy constraints, since fixed power allocation inefficiently consumes energy and limits onboard computing resources. Joint optimization of the DNN partitioning vector, computing frequencies, and transmit powers therefore achieves a better computation--transmission energy tradeoff.

\section{Conclusion}
In this letter, we studied a DNN partitioning based satellite--terrestrial collaborative inference system with multiple LEO satellites and a GS. We formulated an average task completion latency minimization problem by jointly optimizing DNN partitioning, satellite- and GS-side computing resources, bandwidth allocation, and satellite transmit power. To address the coupled discrete and continuous variables, we developed a two-layer optimization algorithm based on closed-form updates, nested bisection searching, and random-restart coordinate descent. Simulation results showed that the gains over AO-GS and random partitioning validate the proposed inner- and outer-layer optimization methods, respectively, while the improvements over binary offloading and fixed resource-allocation schemes demonstrate the benefits of jointly optimizing DNN partitioning and communication-computing resources.

\bibliographystyle{IEEEtran}
\bibliography{references}

@article{leyva2020leo,
  author  = {I. Leyva-Mayorga and B. Soret and M. R{\"o}per
             and D. W{\"u}bben and B. Matthiesen and A. Dekorsy
             and P. Popovski},
  title   = {{LEO} Small-Satellite Constellations for {5G}
             and Beyond-{5G} Communications},
  journal = {IEEE Access},
  volume  = {8},
  pages   = {184955--184964},
  year    = {2020},
  doi     = {10.1109/ACCESS.2020.3029620}
}

@article{leyva2023satellite,
  author  = {I. Leyva-Mayorga and M. Mart{\'i}nez-Gost and M. Moretti
             and A. P{\'e}rez-Neira and M. A. V{\'a}zquez
             and P. Popovski and B. Soret},
  title   = {Satellite Edge Computing for Real-Time and
             Very-High Resolution Earth Observation},
  journal = {IEEE Trans. Commun.},
  volume  = {71},
  number  = {10},
  pages   = {6180--6194},
  month   = oct,
  year    = {2023},
  doi     = {10.1109/TCOMM.2023.3296584}
}

@article{tang2021computation,
  author  = {Q. Tang and Z. Fei and B. Li and Z. Han},
  title   = {Computation Offloading in {LEO} Satellite Networks
             With Hybrid Cloud and Edge Computing},
  journal = {IEEE Internet Things J.},
  volume  = {8},
  number  = {11},
  pages   = {9164--9176},
  month   = jun,
  year    = {2021},
  doi     = {10.1109/JIOT.2021.3056569}
}

@article{denby2019orbital,
  author  = {B. Denby and B. Lucia},
  title   = {Orbital Edge Computing: Machine Inference in Space},
  journal = {IEEE Comput. Archit. Lett.},
  volume  = {18},
  number  = {1},
  pages   = {59--62},
  month   = {Jan.--Jun.},
  year    = {2019},
  doi     = {10.1109/LCA.2019.2907539}
}

@article{diana2024review,
  author  = {L. Diana and P. Dini},
  title   = {Review on Hardware Devices and Software Techniques
             Enabling Neural Network Inference Onboard Satellites},
  journal = {Remote Sens.},
  volume  = {16},
  number  = {21},
  pages   = {3957},
  year    = {2024},
  doi     = {10.3390/rs16213957}
}

@inproceedings{chen2024energy,
  author    = {Y. Chen and Q. Zhang and R. Xing and Y. Li
               and X. Ma and C. Yu and Y. Zhang and A. Zhou
               and S. Wang},
  title     = {Energy-Aware Satellite-Ground Co-Inference via
               Layer-Wise Processing Schedule Optimization},
  booktitle = {Proc. 15th Asia-Pacific Symp. Internetware
               (Internetware '24)},
  address   = {Macau, China},
  pages     = {303--312},
  year      = {2024},
  doi       = {10.1145/3671016.3674811}
}

@article{delprete2025optimizing,
  author  = {{Del Prete}, R. and P. K. Thind and A. Mazzeo
             and M. Whitley and L. Papa and N. Long{\'e}p{\'e}
             and G. Meoni},
  title   = {Optimizing Deep Learning Models for On-Orbit
             Deployment Through Neural Architecture Search},
  journal = {Sci. Rep.},
  volume  = {15},
  number  = {1},
  pages   = {37783},
  month   = oct,
  year    = {2025},
  doi     = {10.1038/s41598-025-21467-8}
}

@article{eshratifar2021jointdnn,
  author  = {A. E. Eshratifar and M. S. Abrishami and M. Pedram},
  title   = {{JointDNN}: An Efficient Training and Inference Engine
             for Intelligent Mobile Cloud Computing Services},
  journal = {IEEE Trans. Mobile Comput.},
  volume  = {20},
  number  = {2},
  pages   = {565--576},
  month   = feb,
  year    = {2021},
  doi     = {10.1109/TMC.2019.2947893}
}

@article{tian2026dynamic,
  author  = {S. Tian and R. Wang and J. Hao and Q. Wu and D. Niyato},
  title   = {Dynamic Collaborative Inference for Multi-Type
             {DNN} Tasks in Space-Ground Networks:
             A {DRL} Approach},
  journal = {IEEE Trans. Cogn. Commun. Netw.},
  volume  = {12},
  pages   = {3265--3282},
  year    = {2026},
  doi     = {10.1109/TCCN.2025.3616114}
}

@inproceedings{chen2023energy,
  author    = {Y. Chen and Q. Zhang and Y. Zhang and X. Ma
               and A. Zhou},
  title     = {Energy and Time-Aware Inference Offloading for
               {DNN}-based Applications in {LEO} Satellites},
  booktitle = {Proc. IEEE 31st Int. Conf. Netw. Protocols (ICNP)},
  address   = {Reykjavik, Iceland},
  pages     = {1--6},
  year      = {2023},
  doi       = {10.1109/ICNP59255.2023.10355644}
}

@inproceedings{liu2025joint,
  author    = {P. Liu and S. Tang and K. Lin and N. Ha
               and X. Wang and Z. Fei},
  title     = {Joint Split Inference Design and Bandwidth Allocation
               for Intelligent Edge Perception Network},
  booktitle = {Proc. IEEE/CIC Int. Conf. Commun. China Workshops
               (ICCC Workshops)},
  address   = {Shanghai, China},
  pages     = {388--392},
  month     = aug,
  year      = {2025},
  doi       = {10.1109/ICCCWorkshops67136.2025.11148170}
}

@article{liu2026toward,
  author  = {P. Liu and Z. Fei and X. Wang and X. Li and W. Yuan
             and Y. Li and C. Hu and D. Niyato},
  title   = {Toward Intelligent Edge Sensing for {ISCC} Network:
             Joint Multi-Tier {DNN} Partitioning and Beamforming
             Design},
  journal = {IEEE Trans. Wireless Commun.},
  volume  = {25},
  pages   = {6774--6789},
  year    = {2026},
  doi     = {10.1109/TWC.2025.3626555}
}

@article{zhong2025joint,
  author  = {L. Zhong and Y. Li and M.-F. Ge and M. Feng and S. Mao},
  title   = {Joint Task Offloading and Resource Allocation for
             {LEO} Satellite-Based Mobile Edge Computing Systems
             With Heterogeneous Task Demands},
  journal = {IEEE Trans. Veh. Technol.},
  volume  = {74},
  number  = {7},
  pages   = {11337--11352},
  month   = jul,
  year    = {2025},
  doi     = {10.1109/TVT.2025.3549119}
}

@article{li2023throughput,
  author  = {J. Li and W. Liang and Y. Li and Z. Xu
             and X. Jia and S. Guo},
  title   = {Throughput Maximization of Delay-Aware {DNN} Inference
             in Edge Computing by Exploring {DNN} Model Partitioning
             and Inference Parallelism},
  journal = {IEEE Trans. Mobile Comput.},
  volume  = {22},
  number  = {5},
  pages   = {3017--3030},
  month   = may,
  year    = {2023},
  doi     = {10.1109/TMC.2021.3125949}
}

@article{razaviyayn2013unified,
  author  = {M. Razaviyayn and M. Hong and Z.-Q. Luo},
  title   = {A Unified Convergence Analysis of Block Successive
             Minimization Methods for Nonsmooth Optimization},
  journal = {SIAM J. Optim.},
  volume  = {23},
  number  = {2},
  pages   = {1126--1153},
  year    = {2013},
  doi     = {10.1137/120891009}
}

@book{boyd2004convex,
  author    = {S. Boyd and L. Vandenberghe},
  title     = {Convex Optimization},
  publisher = {Cambridge University Press},
  address   = {Cambridge, U.K.},
  year      = {2004}
}

@inproceedings{kang2017neurosurgeon,
  author    = {Yiping Kang and Johann Hauswald and Cao Gao
               and Austin Rovinski and Trevor Mudge
               and Jason Mars and Lingjia Tang},
  title     = {Neurosurgeon: Collaborative Intelligence Between
               the Cloud and Mobile Edge},
  booktitle = {Proc. 22nd Int. Conf. Architectural Support for
               Programming Languages and Operating Systems (ASPLOS)},
  pages     = {615--629},
  year      = {2017},
  doi       = {10.1145/3037697.3037698}
}

\end{document}